\documentclass[reprint,amssymb, amsmath,aip,jap]{revtex4-1}

\usepackage{graphicx}
\usepackage{dcolumn}
\usepackage{bm}
\usepackage{graphics}

\newcommand{\xtext}[1]{\mbox{#1}}

\begin{document}

\title{Mechanisms of Stranski-Krastanov Growth}

\author{Arvind Baskaran}
 \email{baskaran@math.uci.edu}
\affiliation{Department of Mathematics, University of California Irvine, California 92697-3875.}

\author{Peter Smereka}
 \altaffiliation[Also at : ]{Michigan Center for Theoretical Physics, University of Michigan, Ann Arbor, Michigan 48109.}
 \email{psmereka@umich.edu}
 \affiliation{Department of Mathematics, University of Michigan, Ann Arbor, Michigan 48109.}

\date{\today}

\begin{abstract}

Stranski-Krastanov (SK) growth is reported experimentally as 
the growth mode that is responsible for the transition to 
three dimensional islands in heteroepitaxial growth. A
kinetic Monte Carlo (KMC) model is proposed that can replicate many of
the experimentally observed features of this growth mode.
Simulations reveal that this model effectively captures the SK transition 
and subsequent growth. Annealing simulations demonstrate that the 
wetting layer formed during SK growth is stable, with entropy 
playing a key role in its stability.
It is shown that this model also captures the apparent critical thickness
that tends to occur at higher deposition rates and
for alloy films (where intermixing is significant).
This work shows that the wetting layer thickness increases with 
increasing temperature whereas
the apparent critical thickness decreases with increasing temperature.
Both of which are in agreement with experiments. 

\end{abstract}

\maketitle

\vspace{-.9in}
\section{Introduction}
Heteroepitaxy is a process 
in which a film is deposited on a substrate of a different material.
The typical examples of film substrate combinations include Ge/Si, InAs/GaAs and InP/GaAs.
The lattice mismatch between the film and substrate can lead 
to the growth of a strained film.
The film often grows in a layer-by-layer fashion until a certain critical thickness,
beyond which 3D islands form through what is known as the Stranski-Krastanov (SK) transition.
In a strained film, the formation of islands is energetically favorable, 
as it reduces the strain energy in the crystal.
The experimental observations on Ge/Si systems seem to show a critical thickness of 
about 3 monolayers (ML) for pure Ge on Si \cite{SK1,SK2}.
Similar observations are made in InAs/GaAs systems \cite{petroff_denbaars}.
These 3D islands show potential to serve as quantum dots and are of great practical importance
\cite{SK1,SK2}.

The instability of a planar film is understood in the context of the Asaro-Tiller-Grinfeld (ATG) 
instability \cite{ATG}.  
However, according to the ATG theory a strained film of any finite thickness is unstable i.e. it predicts
a zero critical thickness.
Currently two fundamentally different theories have been used to reconcile this.
One is based on a kinetic stabilizing mechanism (the notion of an apparent critical thickness)
and the other an energetic mechanism (use of a wetting potential).
 
The concept of an apparent critical thickness relies on the notion
that the instability is present but not observed until the perturbations 
have grown large enough to be perceived 
as islands \cite{svd,tu_tersoff}.
It was further suggested by Tu and Tersoff  \cite{tu_tersoff} that 
intermixing lowers the growth rate of the instability, leading
to a delayed perception of the island formation.
One possible problem with this interpretation is that the region between the islands will 
erode below the original substrate location and this does not appear to be observed in experiments. 

On the other hand, Leonard et al \cite{leonard_pond_petroff} observed that a critical thickness was observed
when depositing InAs on GaAs even at very low fluxes. This suggests an energetic mechanism
might be appropriate; to this end one can introduce a wetting potential, which arises
naturally when assuming that the surface energy varies with height \cite{ORS,SVD,LGV}.
This can give rise to a linearly stable flat film. 
Earlier studies on the wetting interactions date back to the work of Tersoff \cite{TTT}. More
recently these calculations were repeated using density functional theory \cite{BWA}.
It is argued in literature \cite{TTT,BWA} that
approximately 1-3 ML of Ge on Si will be stable.
In both cases this is based on calculating the energy per atom 
after each consecutive layer of Ge is added.
One issue with this work is that the roles of entropy and intermixing, which can be significant 
at the temperatures typical of molecular beam epitaxy, are not included. As a consequence
such an argument cannot predict the temperature dependence of the 
critical thickness observed in experiments. \cite{temperature}.

In this, work we show that the issues discussed above can be well
understood by using an oft-used solid-on-solid atomistic KMC model. The model
uses a bond counting approach for short range interactions 
and a ball and spring model to account for the long range elastic interactions.
This allows us to capture surface energy and elastic energy, 
the basic ingredients present in both approaches outlined above. 
Wetting interactions are put into place using a
species dependent bond strength.
This model not only predicts the existence of a kinetically 
limited apparent critical thickness (as in the work of Tu et al \cite{tu_tersoff}) but also
a true critical thickness for the onset of the instability (not observed in Tu et al \cite{tu_tersoff}).
Entropy is found to play a crucial role in stabilizing the wetting layer in this regime.
The identification of the different stabilizing mechanisms and the role of entropy 
allows us to present a comprehensive theory that explains 
the Stranski-Krastanov transition under different growth conditions.
This approach not only unifies the pre-existing theories but also recovers the 
experimental observations of temperature
dependence and long term stability (to annealing) of the wetting layer.
This is the key contribution in this paper.
The discussion and the model description are presented below for Ge/Si 
but are applicable to heteroepitaxial growth in general.

\section{KMC Model}

The KMC model used in this study was first presented  by Orr et al \cite{Orr} 
and later by Lam et al \cite{LLS}.
An efficient method for solving the model which is used here was presented in our 
previous work \cite{BDS}. 
The model is an atomistic model in which the crystal occupies a semi infinite substrate of Si
on top of which sits a film of deposited material.
The atoms occupy sites on a cubic lattice.

In this work we consider a 1+1 dimensional system of atoms within the solid-on-solid framework (no overhanging atoms).
Each atom in the system is bonded to its nearest (4 possible) and next to nearest (4 possible) neighbors by chemical bonds 
of strength $\gamma$ that depends on the species involved.
There are three possible bonds have strengths $\gamma_{\text{Ge-Ge}}, \gamma_{\text{Ge-Si}}$ and $\gamma_{\text{Si-Si}}$.
These bond energies account for the chemical energy of the system.

The atoms in the system are also connected to their occupied nearest and next to nearest neighbors 
by means of Hookean spring (linear stress strain relations).
This accounts for the long range elastic interactions though linear elasticity.
The lateral and vertical springs connecting nearest neighbors have spring constant $k_L$
and the diagonal springs connecting the next to nearest neighbors have strength $k_D$.
The spring constants are chosen so that they satisfy $k_D = k_L/2$ corresponding to the
case of isotropic linear elasticity \cite{BDS,schulze_smereka}.
The natural bond lengths or lattice spacings of the atoms are denoted by $a_{ss}, a_{gs}$ and $ a_{gg}$
corresponding to Si-Si, Ge-Ge and Ge-Si bonds.
They are not equal in general and are chosen to reflect the Ge/Si system with a 4\% misfit \cite{PARAMETERS}.

The hopping rate of a surface atom is given by
\begin{equation}
R_\ell = R_0 \exp \left( \displaystyle{ \frac{ \Delta E + E_0 }{k_B T} }\right)
\end{equation}
where $\Delta E$ is the total change in energy when the atom is removed from the
system and $k_B T$ is the thermal energy.
The parameters $R_0$ and $E_0$ are chosen to give physically relevant hopping rates \cite{PARAMETERS,hull}.
This choice of Arrhenius-like rates satisfies detailed balance.
This has two parts the chemical energy (bond counting) and the elastic energy (springs) of the system. 
\begin{equation}
\Delta E = \Delta E_{chem} + \Delta E_{elas}
\end{equation}

The chemical energy of the system is accounted for by bond counting.
It is  simply the sum total of the energy in the bonds that need to be broken to remove the 
atom from the system.
For example in the case of a pure Si  system the change in chemical energy of the system
is simply $ - N \gamma_{\text{Si-Si}} $ where  $N$ is the number of occupied neighbors.
The hopping rate in this case is simply

\begin{equation}
R_\ell = R_0 \exp \left( \displaystyle{ \frac{ -N \gamma_{\text{Si-Si}}+ E_0 }{k_B T}} \right).
\end{equation}

In the more complex case of a mixture of Ge and Si the change in chemical energy 
is calculated by adding the energies associated with each bond that is broken to remove the atom.
The general form is written as
\begin{equation}
R_\ell = R_0 \exp \left(\displaystyle{ \frac{\Delta E_{chem} + \Delta E_{elas} + E_0}{k_B T}}
\right) 
\end{equation}
where $\Delta E_{chem}$ is the change in chemical energy (bond energy), 
$\Delta E_{elas}$ the change in elastic energy when the atom is removed from the system.

 Not surprisingly a majority of the events during the course of a KMC simulation
are adatom ($N \leq 3$)hops. This means the total computational time needed significantly depends
on the treatment of the adatoms. To increase the simulation speed, the elastic 
contribution and the species type are ignored when evaluating the adatom hopping rates.
Naturally, one should be concerned when making an approximation such as this;
but nevertheless extensive tests revealed that little loss of fidelity was observed
while the code was 3 to 5 times faster\cite{BDS}. For this reason, the results in this
paper use the following rates
\begin{equation}
R_\ell = \left\{
\begin{array}{cc}
 R_0\exp \left({\frac{-3 \gamma_{\text{Ge-Si}} + E_0}{k_B T}}\right) & N \leq 3\\ \\
 R_0 \exp \left(\displaystyle{ \frac{\Delta E_{chem} + \Delta E_{elas} + E_0}{k_B T}}
\right) & N > 3 
\end{array} \right.
\end{equation} 

The KMC simulation of this model is computationally very expensive 
since the elastic displacement field must be updated repeatedly. For the simulations presented in this
study over $10^9$ updates are typically needed. Therefore it is particularly crucial to perform
this operation with great efficiency.   This is achieved by using both global and local updates
of the displacement field. Local updates are computed using the expanding box method, and
when this local approach fails global updates are used \cite{schulze_smereka,BDS}.
The global approach is based on a multigrid algorithm combined with an artificial
boundary condition (to incorporate the semi-infinite substrate) \cite{RS1,RS2}. 
Our algorithm involves a reduced-rejection KMC, 
based on the use of upper-bounds for the rates and a rejection to compensate.
The upper bounds are obtained from local (computationally inexpensive) overestimates of
the change in elastic energies ($\Delta E_{elas}$). They
are reasonably sharp, leading to rejection rates of under 5\% \cite{schulze_smereka, BDS}. 
The reader is directed to our previous  work \cite{BDS} for a comprehensive 
exposition of the algorithm and underlying theory.

\begin{figure*}[!ht]
\includegraphics[width=6.6in]{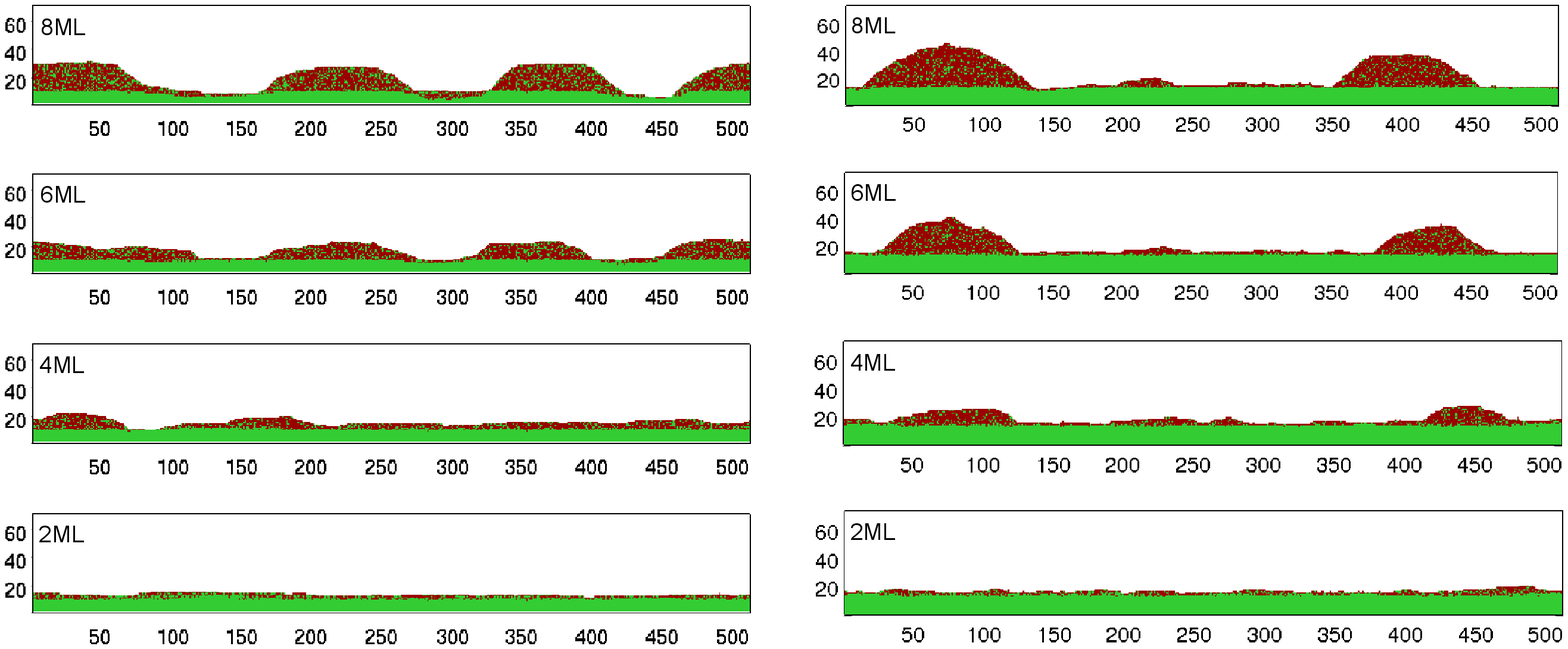}
\caption{
Shows the snapshots of the growth of pure Ge on flat Si substrate at 600K under a deposition flux of
0.8ML/s. The plots show the film at 2,4,6 and 8ML of deposition. The simulation was performed on a periodic domain of 512 atoms in the horizontal direction.\\
{Left} - Bond strengths are equal:  $\gamma_{Si-Si}=\gamma_{Si-Ge} =\gamma_{Ge-Ge} = 0.37 \text{ eV}$ \\
{Right} - Bond strengths are unequal: $\gamma_{Si-Si}=0.37 , \gamma_{Si-Ge} =0.355 \text{ and }\gamma_{Ge-Ge} = 0.34 \text{ eV}$ \\
 }
\label{fig:diff_bonds_comparison}
\end{figure*}

\begin{figure}[!ht]
\includegraphics[width=3.3in]{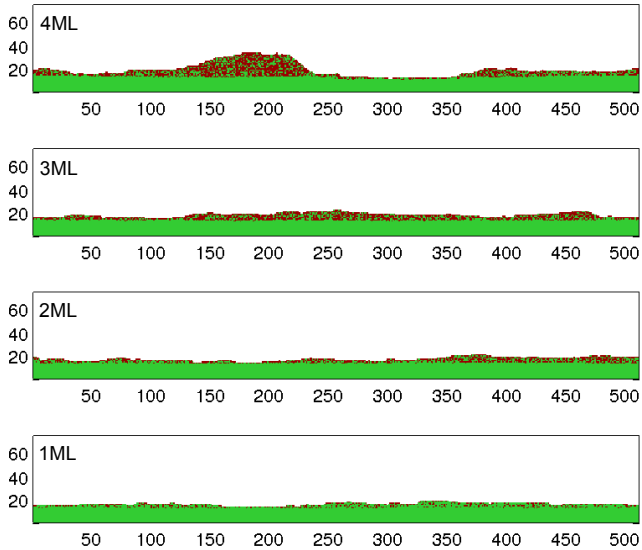}

\caption{
Shows the equilibrium configuration obtained upon annealing a flat film of 1ML, 2ML, 3ML and 4ML  of Ge on 
a flat Si substrate
at a temperature of  600K. The simulation was performed on a periodic domain of 512 atoms in the horizontal direction.
}
\label{fig:diff_bonds_diff_thickness}
\end{figure}

\begin{figure}[!ht]
\includegraphics[width=3.3in]{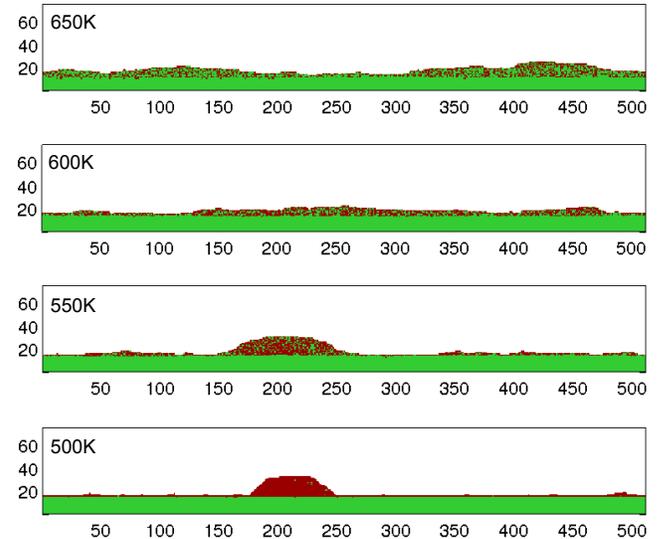}

\caption{
Shows the equilibrium configuration obtained upon annealing a flat film of 3ML of Ge on 
a flat Si substrate
at a temperature of  500K, 550K, 600K and 650K. The simulation was performed on a periodic domain of 512 atoms in the horizontal direction.
}
\label{fig:diff_bonds_diff_temp}
\end{figure}

\section{Stranski-Krastanov Growth Mode}
The above model sets stage for our discussion on the SK growth mode.
  Here we choose 
$\gamma_{\rm Ge-Ge} = 0.34 eV, \gamma_{\rm Ge-Si} = 0.355 eV$ and $\gamma_{\rm Si-Si} = 0.37 eV$
 \cite{BDS,PARAMETERS}.
The system is maintained at a temperature of 600K  and atoms are deposited 
 at a rate of 0.8 monolayers per second on a periodic domain of size 512 atoms.
Four snapshots of the simulations are shown in Fig. \ref{fig:diff_bonds_comparison}(right). We observe
layer-by-layer growth up to approximately 3 ML after which 
islands nucleate indicating the critical thickness is approximately 3 ML.
Subsequent growth reveals islands growing on a Germanium rich layer
of approximately 3 ML indicating  presence of a stable wetting layer.
We remark that there exists small trenches near the periphery of the islands.
This growth mode persists even after 10 ML of deposition.
These observations are consistent with SK growth.

To further validate our claim that this is indeed SK growth,
we perform the following annealing experiments.
We consider an initial condition of $N$ ML of Ge placed on
a substrate of Si which is then annealed for approximately 14 seconds
(this is about $10^{10}$ KMC steps at 600 K). These simulations 
(see Fig. \ref{fig:diff_bonds_diff_thickness})  indicate that
Ge films that are  $\le 3$ ML are stable \cite{metastable} but at 4 ML
islands will form. From this we conclude that it is reasonable to surmise
that this KMC model has a true critical thickness and 
wetting layer thickness of 3 ML.
This is indeed the observation made in experiments  \cite{leonard_pond_petroff}:
that a sharp critical thickness is observed even at very low 
deposition rates. The low deposition rate  ensures that 
the film is close to equilibrium
and these observations are not the result of kinetically
limited processes.  

It should be pointed out
that these stable films are stressed and so it is surprising, that islands do not form. We claim the reason that
islands do not form is due to entropy. This is to say
the energetic advantage to forming an island cannot overcome entropy.
To validate this assertion we preformed a similar series of
annealing simulations, this time we annealed 3 ML of
Ge on Si at different temperatures (see Fig. \ref{fig:diff_bonds_diff_temp}). These
simulations show that if  $T \le 550K$ islands will form.
On the other if $T \ge 600K$ islands do not form. Clearly if
the temperature is too large entropy will prevent islands
from forming even with the wetting layer being stressed.
Also the true critical thickness increases with temperature as
observed in experiments done at low fluxes \cite{temperature}.

 It was first observed experimentally by Mo et al \cite{MO} that metastable 3D
islands exist in the Ge/Si system well before the formation
of stable 3D islands. This was also confirmed by Yam et al \cite{YAM} using RHEED and AFM techniques.
They conclude that the film loses stability before stable 3D islands can form
by instead producing very small metastable 3D islands by strain relaxation.
Similar observations were made by Floro et al \cite{FLORO2} for alloy films.

This is consistent with our simulations, 
the reader is directed to  the  right hand side of Fig. \ref{fig:diff_bonds_comparison} which
shows tiny islands after 2 ML of deposition.
Even though these small islands lower the strain energy, the
energetic advantage of their formation is not quite enough to overcome entropy.
Therefore,  they are only metastable. Only when enough material
has been deposited can islands form that are big enough to yield a reduction
in strain energy that can overcome entropic effects. The annealing results, for a 2ML film 
shown in Fig. \ref{fig:diff_bonds_diff_thickness}, reveal a smoother film with fewer
tiny islands which is consistent with the assertion that the tiny islands in the growing
film are indeed metastable.

Finally, we mention the last crucial ingredient contained in
this model needed to capture SK growth
is allowing the bonds strengths to be different.
This can be understood  by performing 
a simulation, the same as above, but with identical bond strengths:
$ \gamma_{Si-Si} = \gamma_{Ge-Si} = \gamma_{Ge-Ge} = 0.37 \xtext{ eV} $.
This simulation produces an apparent critical layer but
as the film grows trenches form between the islands 
destroying the wetting layer. The result of growth 
are shown in Fig.  \ref{fig:diff_bonds_comparison} (left). The reason the trenches form
in this case and not the other is twofold.  Because all the bonds
strengths are the same, there is a lower energy barrier to intermixing.
Surprisingly, intermixing actually increases the total elastic
energy which further drives the ATG instability resulting in the
trench formation. Table \ref{energy_table} shows that after 8 ML
of deposition the equal bond case has considerably more elastic
energy than the case with different bond strengths.
Thus elastic effects can make segregation to be energetically preferred
since intermixing increases the elastic energy.
This suggests that the intermixing is in fact entropically preferred and the
use of different bond strengths provides the much needed energy driven
pathway to segregation ultimately leading to the SK growth mode.

\begin{table}[ht]
\vspace{-0.2in}
\caption{Total elastic energy of the crystal configuration obtained when 8ML of pure Ge on Si deposited at 0.8 ML/s
at 600K on a periodic domain of 512 atoms}
\begin{center}
  \begin{tabular}{|l|c|} \hline
  Equal Bond Strengths     & 795.5602  $k_BT$         \\[0.03in]
  \hline
  Different Bond Strengths  & 637.4619  $k_BT$\\[0.03in]
  \hline
  \end{tabular}
\end{center}
\label{energy_table}
\vspace{-0.3in}
\end{table}

\section{The Apparent Critical Thickness}
Intermixing plays a more significant role when 
depositing Si-Ge alloys as compared to
pure Ge on Si.  In this case,  we initially observe
an apparent critical thickness much like that reported
by Tu and Tersoff\cite{tu_tersoff}.
We have compiled a series of simulations where
$\xtext{Ge}_{x}\xtext{Si}_{1-x}$ is deposited  on
Si with  $\gamma_{\rm Ge-Ge} = 0.34 eV, \gamma_{\rm Ge-Si} = 0.355 eV$ and $\gamma_{\rm Si-Si} = 0.37 eV$.

 In order to quantify our observations, we introduce the quantity, $\cal F$, which is taken
to be the amplitude of the dominant mode of the discrete Fourier transform of the surface height function.

\begin{figure}[!htt] 
\begin{center}
\includegraphics[width=3.2in]{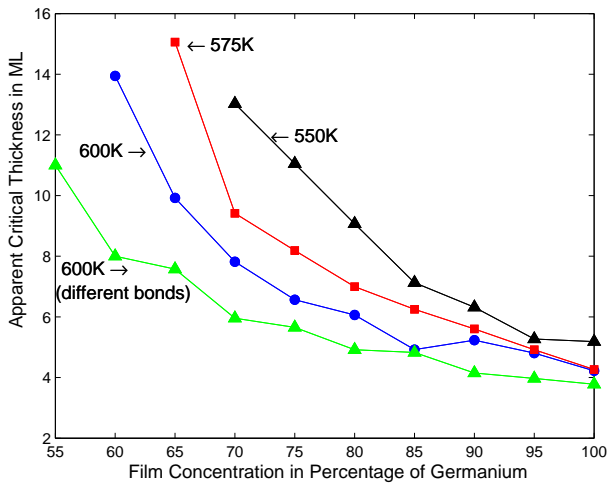}

\end{center}
\caption{
Shows the apparent critical thickness as a function of concentration of Ge ($x$) for deposition of
$\xtext{Ge}_{x}\xtext{Si}_{1-x}$ on Si.
The results were averaged over 4 independent runs. 
The simulation was performed on a periodic domain of 1024 atoms in the horizontal direction.
 }
\label{fig:critical_thickness}
\end{figure}

 To perform a quantitative analysis of the dependence of the apparent critical thickness on temperature 
 and film concentration,
we define the apparent critical thickness to be the average thickness of deposited film, 
where the quantity $\cal F$, 
is 1000 (roughly 5 times the magnitude of a kinetically roughened film ( $\approx 200$)).
This definition is similar to the one used by Tu and Tersoff \cite{tu_tersoff}.
The Fig. \ref{fig:critical_thickness} shows the apparent critical thickness as a function of concentration of Ge in the 
deposited material.
These results were obtained by ensemble averaging the data obtained on a periodic domain of 1024 atoms over 4 independent runs. 

\begin{figure}[!ht]
\includegraphics[width=3.3in]{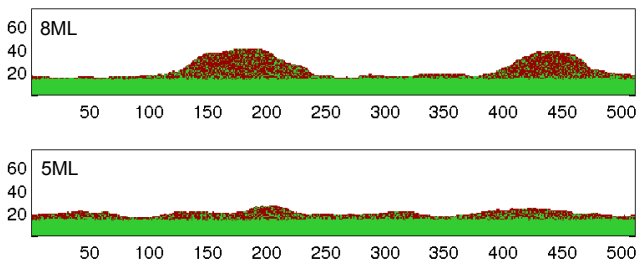}

\caption{
This figure shows a simulation where an alloy ($\xtext{Ge}_{0.8}\xtext{Si}_{0.2}$) 
is deposited at 0.8 ML/sec on Si at 600K with 
bond strengths $\gamma_{Ge-Ge} = 0.34$, $\gamma_{Ge-Si} = 0.355$, and $\gamma_{Si-Si} =0.37eV$.
The apparent critical thickness is 5 ML however 
further growth results in wetting layer much smaller than 5 ML.
The simulation was performed on a periodic domain of 512 atoms in the horizontal direction.
}
\label{fig:apparent}
\end{figure}

These simulations
show that apparent critical thickness
can be as large as 11 ML for $x=.55$. 
They also demonstrate
(see Fig. \ref{fig:critical_thickness}) 
that the apparent critical thickness increases
with decreasing $x$ and decreasing temperature. 
This is also in agreement with experiments \cite{chinese, FLORO2}.
This behavior is understood by noting that as the temperature increases the entropy driven 
intermixing dilutes the film further slowing the growth rate of the instability.
The effect is found to be more pronounced in the case of dilute deposition flux.
Due to the computational cost the temperature dependence is only shown with equal bond strengths.
Preliminary tests indicate a similar trend persists with unequal bond strengths.
In addition our simulation show that
subsequent growth sees this
apparent critical layer disappearing  with its contents
being incorporated into the islands leaving behind
a wetting layer of about 3 ML. This behavior was
observed in experiments \cite{floro}.
Fig. \ref{fig:apparent} shows an example of this in a simulation of $\text{Ge}_{0.8} \text{Si}_{0.2}$
growth on Si.
Here an apparent critical thickness of about 5 ML is observed (no islands at 5 ML growth).
However at 8ML of growth well separated islands are formed where some of the material 
from the critical layer (first 5 layers) is incorporated into the islands.
At 8ML of growth a wetting layer much smaller than 5ML is observed between the islands. 

\begin{table}[!ht]
\caption{The apparent critical thickness as function of the deposition rate for Ge on Si
at 600K on a periodic domain of 512 atoms for the case of different bond energies}
\begin{center}
  \begin{tabular}{|l|c|c|c|c|c|c|} \hline
  \multicolumn{7}{|c|}{Apparent Critical Thickness as Function of Deposition Flux} \\[0.03in]
  \hline
  Deposition Flux in ML/s       &0.2&0.4&0.8&1.6&3.2&6.4     \\[0.03in]
  \hline
  Critical Thickness in ML      &3.66&3.79&4.22&4.73&4.74&5.40 \\[0.03in]
  \hline
  \end{tabular}
\end{center}
\label{flux_table}
\end{table}

In addition simulations also indicate a weak dependence of the 
apparent critical thickness at low flux regimes with the apparent critical thickness
decreasing with the flux (toward the true critical thickness).
Table \ref{flux_table} shows the apparent critical thickness of pure Ge on Si as a function of deposition flux.
The trends are consistent with experiments \cite{SK2}.

\section{Different Mechanisms involved in Stranski-Krastanov Growth}
Based on the above results, we conclude there are primarily three different mechanisms responsible
for the phenomena associated with Stranski-Krastanov growth.
These are:
\begin{itemize}
\item Apparent Critical Thickness
\item Wetting potentials
\item Entropic stability
\end{itemize}
\subsection{Apparent Critical Thickness }
The first case we shall address is the mechanism related to
the notion of an apparent critical thickness\cite{svd} which tends to arise
when the growth rate associated with the ATG instability and the deposition rate are comparable.
We observe that kinetic roughening and the adatom diffusion lead to enough intermixing
to reduce the strength of the elastic instability and thereby
increasing the apparent critical thickness, also in agreement with Tu  and Tersoff\cite{tu_tersoff}.
Our simulations suggest that strained film growth in this regime has three stages.
\begin{itemize}
\item{\bf Stage I :}
{\em The first stage occurs when the film exhibits a planar layer-by-layer growth with three regions
being observed: the substrate, an intermixing layer, and the film.
The concentration of the Ge increases as one moves from substrate to the surface.
At the latter potion of this stage the film will have small islands (metastable) and/or ripples.}
\item{\bf Stage II :}
The second stage begins when the film thickness exceeds an apparent
critical thickness marked by the growth of the small metastable islands into
stable quantum dots. 
\item{\bf Stage III :}
The third and final stage is characterized by well separated islands on top of stable wetting layer,
whose thickness will be less than the apparent critical thickness with
the excess material being incorporated into the stable islands. The wetting layer is
only observed when different bond strengths are used.  In the case of equal bond strengths, 
what might have been a wetting layer is, instead, a layer that 
is eroded down to the substrate in the valleys between the islands.
\end{itemize}

This three stage process is consistent with a number of experiments \cite{floro}.
It is found that the difference is surface energies between the film and the substrate is a crucial 
ingredient in the stabilization of the wetting layer, leading to our next mechanism.

\subsection{Wetting Potentials}
The introduction of wetting potential\cite{ORS,SVD,LGV}
allows one to include the possibility that the surface
energy may change with film thickness 
due to film/substrate interactions. When chosen 
appropriately, the wetting potential can stabilize a thin
film in the presence of strain.
In our model these effect are naturally incorporated by the use
of different bond strengths.  The difference in bond strengths 
leads to the difference in surface energies between the two species.
For our choice (physically relevant) the Ge surface energy is lower 
and it is cheaper to have a Ge rich surface.

When the temperature is low and as a consequence, entropy is not dominant we observe a wetting layer of 
one atomic layer thickness (see Fig. \ref{fig:diff_bonds_diff_temp}, T= 500K).
Even though this layer is strained it still energetically favorable
because the Ge surface has less surface energy than a Si surface. However, at low temperatures
the thicker film would not be stable because of strain relaxation,
indeed if we again look at Fig. \ref{fig:diff_bonds_diff_temp} (T= 500K) we infer that film
of thickness 3ML is not stable. However, as the temperature is increased this scenario
changes due entropic contributions which provide a stabilizing mechanism discussed below.

\subsection{Entropic Stabilization}
The last mechanism we will discuss results from entropic contributions. The effects
of entropy can allow a film that would be energetically unstable at low temperature
to become stable at sufficiently high temperatures. Such a film would be linearly
unstable by the Asaro-Tiller-Grinfeld instability  (even accounting for the wetting potential).
However, entropy can prevent the surface height perturbations from growing into islands.

In the regime, when films are grown at ultra low fluxes, entropic effects may be
the  dominant mechanism.  The film,  grows in a layer-by-layer fashion and the strain
energy increases. Only, when the energetic advantage of island formation is enough 
to overcome entropy,  will 3D islands form. The so formed islands sit on top
of a wetting layer.  The extreme case of zero flux is explored 
using annealing simulations as presented in
Fig. \ref{fig:diff_bonds_diff_thickness} and  Fig. \ref{fig:diff_bonds_diff_temp}.
Our conclusions on this regime are consistent
with experiments at ultra low fluxes \cite{petroff_denbaars}.

\section{Summary}

 The phenomena of SK growth is an interplay involving the aforementioned  mechanisms,
but overall  SK growth can be understood as the competition between energy and entropy.
Accounting for entropy allows one to bridge the two existing theories of 
apparent critical thickness and wetting potentials to formulate a comprehensive theory 
that also explains regimes not accounted for by the individual theories.

In conclusion, we have presented a simple KMC model
that is able to capture many aspects of heteroepitaxial
growth including wetting layers  and
apparent critical layers.
The distinction between the apparent critical thickness 
and wetting layers thickness has been clarified.
Finally the importance of the roles of entropy 
and intermixing has been identified.


\section*{Acknowledgements}
The authors gratefully acknowledge helpful conversations with Joanna Mirecki-Millunchick,
Len Sander, Vivek Shenoy, and Jerry Tersoff. This research was supported, in part,
with grants DMS-0553487, DMS-0509124, DMS-0810113,
DMS-0854870, and DMS-1115252 from the National Science Foundation.
One of the authors AB was supported in part by a postdoctoral fellowship from the
Institute for Pure and Applied Mathematics
and University of California Irvine.  The authors also thank Bennet Fauber for help with
computing resources.


\begin{thebibliography}{99}
\bibitem{metastable}More likely the film is only metastable on very large time scales as discussed by Tersoff\cite{TTT}.
\end{thebibliography}
\end{document}